\documentstyle[aps,prb,psfig,twocolumn]{revtex}
\begin{document}
\preprint{IASSNS-HEP-97/33}
\draft
\twocolumn[\hsize\textwidth\columnwidth\hsize\csname @twocolumnfalse\endcsname
\title{A Creation Operator for Spinons in One Dimension}
\author{ J.C.\ Talstra$^1$ }
\address{ $^1$ James Franck Institute, Univ. of Chicago, 5640 S. Ellis
Ave, Chicago, IL 60637}
\author{S.P.\ Strong$^2$}
\address{$^2$ Institute for Advanced Study, Olden Lane, Princeton, NJ
08540}
\date{\today}
\maketitle
\begin{abstract}
We propose a definition for a creation operator
for the spinon, the fractional statistics
elementary excitation of the Haldane-Shastry model,
and give numerical and analytical evidence
that our operator creates a single spinon
with nearly unit amplitude in the ISE model.
We then discuss how the operator is useful
in more general contexts such as 
studying the underlying spinons
of other spin chain models, like the XXX- and XY-model, and 
of the one dimensional Hubbard model.
\end{abstract}
%\pacs{}
]
% body of paper here

One of the deepest open problems in condensed matter
physics today is that of understanding fractional statistics,
that is understanding field theories whose underlying excitations
do not obey either Bose or Fermi statistics.  
Such possibilities
(fractional statistics) for point particles
are forbidden in three dimensions;
however, in two dimensions, 
braid statistics can occur and
anyons \cite{anyon} are allowed,
while, in one dimension,
there exist parafermion operators
\cite{parafermions} which also do not obey Fermi
or Bose statistics.
Such fractional statistics objects may 
be important for describing the
low energy excitations of condensed matter systems
if these systems are categorized by fixed points whose
effective, low energy Hamiltonians are
one or two dimensional.

The Heisenberg model with
inverse squared exchange (ISE) \cite{HS,Suth71}
($H = \sum_{ i,j } (\frac{N}{\pi} \sin \pi \frac{i-j}{N})^{-2}
{\bf S}_i \cdot {\bf S}_j$)
is believed to provide a model whose 
excitations are completely non-interacting
fractional statistics particles
\cite{yangian,Hald94} referred to as
{\it spinons}.  The statistics of these
excitations are most naturally discussed in terms
of an alternative definition,
proposed by Haldane
\cite{haldfrac}, in which a particle's statistics
is defined in terms of its effect on the 
size of Hilbert space for other particles.
For example, on the lattice, each fermion added to a system
decreases the number of states in the Hilbert space
for other fermions of the same type by one, while each added
boson leaves the size of the available Hilbert
space unchanged.  Fractional exclusion statistics 
occurs for operators which reduce the Hilbert space
by a non-integer amount, i.e. at a fraction of
the rate for fermions.  This definition
has been shown to agree with
the definition of statistics
through particle interchange
in the case of
the quasiparticles of the Laughlin state
(when the cut-offs in the theory are treated
carefully) \cite{haldfrac,laughlinwvfnc,wu}. 
In that case, the change in the phase of the 
wavefunction resulting from quasihole interchange 
is the same fraction of $\pi$ 
as the rate of Hilbert space reduction due to
the creation of quasiholes,
relative to the rate of reduction due to 
the creation of fermions.

For the exclusion approach, lattice models
are more natural since
the number of states in the Hilbert space
is easily defined, and, indeed, Haldane
introduced his definition 
partially %
motivated
by results for the ISE spin chain model:
the
spectrum of the %
ISE spin chain model
can be described with a set of non-interacting
single particle energies, and a set of rules
for the occupation of the single particle 
levels %
which embody the Yangian symmetry of the
model \cite{yangian}.  The effect of these
rules is that 
the creation of spinons eliminates allowed
spinon states (of any polarization)
from the Hilbert space half
as fast as the creation of fermions eliminates
allowed fermion  single particle states
from a Hilbert space of fermionic  
states \cite{twokindsnote}.
The spinons are thus half-fermions
or semions and are {\it completely
free}.  
This property makes the ISE model somewhat
more natural for studying exclusion
statistics than the Laughlin states
since the quasiholes are not ideal
fractional exclusion statistics particles
\cite{chetan}.

In light of this, it would
be very useful to have some method of treating
the underlying spinons in that and related models
beyond the first quantized formalism; one would like
to introduce 
spinon creation and annihilation operators.
The main purpose of this
work is to introduce an operator which has
essentially unit overlap with the spinon
creation operator in the
ISE model, and which, we conjecture, can be
interpreted as a spinon creation operator
much more broadly than this.

The operator we now propose was motivated chiefly 
by the behavior of the large $U$ wavefunction
of the Hubbard model \cite{OgataShiba},
which factorizes into spin and charge parts
which are linked only by a backflow condition
\cite{OgataShiba}.
Except for the backflow condition,
the charge part is described by free spinless
fermion wavefunction and the spin part 
by a wavefunction of the one
dimensional Heisenberg model
($H = \sum_{i} {\bf S}_i \cdot {\bf S}_{i+1}$)
where the role of the sites is played by the electrons.
The creation
operator for an up spin electron in that model
involves
the creation of a spinless fermion in the
charge wavefunction, the backflow and
the insertion of a site with an up spin into
the Heisenberg model wavefunction.  Likewise,
the bosonization approach to the Hubbard model
\cite{hubb:ll} exhibits a similar spin charge
decoupling, with the electron creation operator
effectively factorizing into a product of
exponentials in spin and charge density
eigenexcitations.  
This suggests the identification
of the spin insertion with the spin boson
part of the electron operator.
Further, the spin operator
is identical to the primary field of the
$c=1$ conformal field theory studied in
\cite{ludwig}, and the Fourier models of this operator
are also recognizable as a semionic parafermion.
Thus, it has the essential
properties expected of a spinon operator,
being both semionic and
providing a natural representation of the Yangian
\cite{ludwig}.

We therefore conjecture that the
local up-spinon creation operator
might be given approximately by
an operator,
$O^{\dagger}(x)$,
which inserts an additional up-spin
into a one-dimensional spin chain
after site $x$ but before site $x+1$. 
Peculiarly, this  operator, in the process
of creating a spinon,
would have to change the length
of the spin chain it acted on from $N$ to $N+1$.
Some such an exotic effect is clearly necessary
in defining a single spinon creation
operator because of the fractional
exclusion statistics of the spinon; an operator 
which creates a {\em single} spinon must involve
the rearrangement of the Hilbert space
different from what occurs for fermions or bosons
and hence the need to {\it change the number of
sites in the model in order to create a single spinon}.
We have discussed this feature of exclusion
statistics before in a 
slightly different context involving spin chains \cite{usODLRO},
and it also occurs in the case of Laughlin wavefunctions
restricted, for example, to the sphere where the 
creation operator involves an addition of allowed states 
to the (appropriately cut-off) Hilbert space.

If we accept this unusual property
as an unavoidable complication for fractional
statistics operators, the 
evidence is there to support the proposal
that the spin insertion creates a single
spinon 
with the same polarization.                               %
Let us first study this conjecture in the context of the ISE model.
In complete analogy with the insertion of two spins in a singlet 
\cite{usODLRO},
we can construct the state obtained by inserting
a single spin. To fix notation:
the unnormalized 
groundstate wavefunction on $N$ (even) sites of the ISE model in a basis 
labeled by the positions of the down-spins
$\left\{ n_1,\ldots ,n_M \right\} ,\,  M=\frac{N}{2}$ is given by:
\begin{equation}
\Psi^{(0)}_N\left(n_1,\ldots,n_{\frac{N}{2}}\right) = \prod_{i=1}^{\frac{N}{2}}
(-)^{n_i}
\prod_{i<j}^{\frac{N}{2}} \sin^2\left(\frac{n_i -n_j}{N}\pi\right).
\label{eq:wfgs}
\end{equation}
In a basis of local spins
$ \{ |\sigma_1\cdots \sigma_N\rangle\}$, $\sigma_i = \pm \frac{1}{2}$ this
reduces to \cite{Hald91}:
\begin{equation}
\Psi^{(0)}_{N}\{\sigma_1\cdots\sigma_N\} = 
\prod_{i<j}^{N}(z_i-z_j)^{\delta_{\sigma_i ,\sigma_j}} 
e^{\frac{\pi i}{2}\rm{sgn}(\sigma_i - \sigma_j)}.
\label{GS}
\end{equation}
For the even $N$-site ISE groundstate $\{z_i\}\equiv C_N =
\{ e^{\frac{2 \pi i
n}{N}};n=1\ldots N \}$.
The action of $O^\dagger (x=N)$ on this state is, in the language of
eq.\ (\ref{GS}), to add an extra spin $\sigma_{N+1}=\uparrow$, and
leave the other spins alone (with trivial generalization to other
values of the insertion-site $x$).

\begin{figure}[htb]
\centerline{\psfig{file=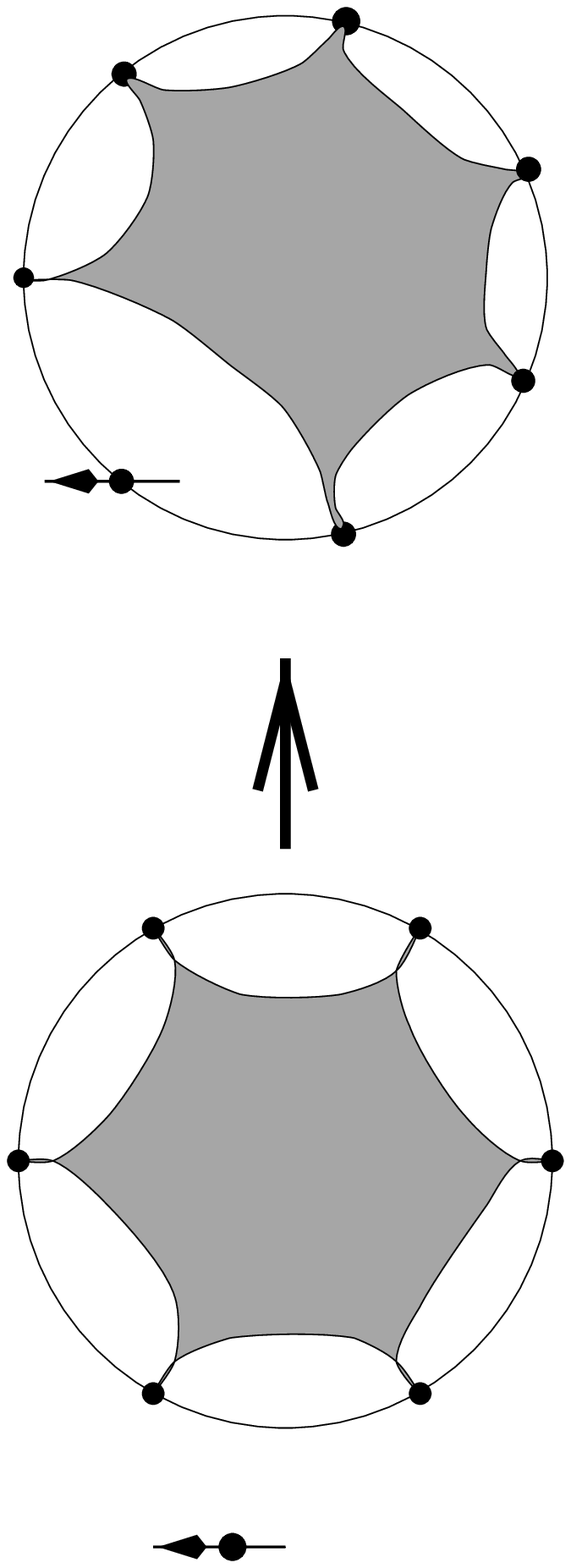,width=\hsize,angle=-90}}
\caption{Graphical illustration of the deformation of the spin
insertion to a spinon singlet insertion. The gray area is the
groundstate wavefunction, or an adiabatic deformation thereof.}
\label{trampofig}
\end{figure}

In order to answer the question, to what extent this state $O^\dagger
(x)|\Psi^{(0)}_N\rangle$ simulates a true 1-spinon state, let us
recall what the latter looks like \cite{Hald91,TH94}. Since the
ISE-Hamiltonian commutes with momentum, the 1-spinon eigenstates
(which exist only for an {\em odd} number of sites) have
definite momentum. However these states can be coherently superimposed
(Fourier transformed) into states in which the single spinon is
completely localized on a site, say $x$, which is no longer an
eigenstate of $H_{\rm ISE}$. Such an unnormalized state is known to be
\cite{Hald91,TH94} of  the form
\begin{eqnarray}
\lefteqn{\Psi^{(1)}_{N+1}(n_1\ldots n_{\frac{N}{2}}|x)=}\hspace{3.5em}
\nonumber\\
&&\prod_{i=1}^{\frac{N}{2}}(-)^{n_i}\sin(\frac{n_i-x}{N+1}\pi) 
\prod_{i<j}^{\frac{N}{2}}\sin^2(\frac{n_i-n_j}{N+1}\pi),
\label{onespinon}
\end{eqnarray}
in the language of down-spin co-ordinates $\{ n_i \}$. In terms of
spin co-ordinates $\{ \sigma_{1}\ldots\sigma_{N+1} \}$ it looks
exactly like eq.\ (\ref{GS})\cite{ftnote1}:
\begin{equation}
\Psi^{(1)}_{N+1}=
\prod_{i<j;\, i,j\neq x}^{N+1}(z_i-z_j)^{\delta_{\sigma_i
\sigma_j}} 
e^{\frac{\pi i}{2} {\rm sgn}(\sigma_i - \sigma_j)},
\label{onespsp}
\end{equation}
but $\sigma_x$ is excluded from the product (since $\sigma_x$ is fixed
to be {\em up}), and the set $\{ z_i \} \equiv C_{N+1} = \{
e^{\frac{2\pi i}{N+1}};\, n=1\ldots N+1\} / \{ z^{\frac{2\pi i
    x}{N+1}} \}$. Notice that in the $\{\sigma_i\}$-basis the
quasihole-factor in (\ref{onespinon}) goes away. Our conjecture is now
that $|\Psi^{(1)}_{N+1}(x)\rangle \simeq O^\dagger
(x)|\Psi^{(0)}_N\rangle$. Numerically it turns out that the two states
have an excellent overlap: $\frac{\langle \Psi^{(1)}_{N+1}(x)|
  O^\dagger (x)|\Psi^{(0)}_N\rangle}{ \parallel\Psi^{(1)}_{N+1}(x )
  \parallel \cdot\parallel \Psi^{(0)}_N\parallel} =0.984 + {\cal
  O}(N^{-\frac{3}{2}})$. This is unsurprising, since
$|\Psi^{(1)}_{N+1}(x )\rangle$ is obtained from $O^\dagger (x
)|\Psi^{(0)}_N\rangle $ by adiabatically deforming the set of $N$
$z_r$'s: $e^{\frac{2\pi i}{N}\cdot r}\rightarrow e^{\frac{2\pi
    i}{N+1}\cdot r}$, and leaving the single spin(on) at site $x$
alone, see Fig.\ \ref{trampofig}. Although this is an uncontrolled
approximation, it seems justified by the high numerical overlap.

Naively one might expect from eq.\ (\ref{onespinon}) there
to be $(N+1)$ 1-spinon states; there is, however, a reduction, because
only $\frac{N+2}{2}$ of these are linearly independent. This can be
seen by reconstructing the momentum eigenstates, through Fourier
transforming eq. (\ref{onespinon}) in the spinon co-ordinate
$x$. This transform only has support in the range $[
-\frac{\pi}{2},\frac{\pi}{2} ]$ ($N=4m$) or
$[\frac{\pi}{2},\frac{3\pi}{2}]$ ($N=4m+2$). One notices that the
states thus obtained are indeed eigenstates, since (1) the set $\{
\Psi^{(1)}_{N+1}(x )|x=0\ldots N\} $ spans the 1-spinon subspace \cite{Hald91},
and (2) there is only {\em one} up-spinon eigenstate for a fixed
value of momentum.

As stated before, spinons in eigenstates are not localized but have a
momentum so that we must investigate to what extent our
spin insertion can be connected with spinon creation in momentum space.
Since the number of sites in the chains
before and after the action of the spinon
operator is different, the definition
of the spinon creation operator of fixed momentum
is not trivial and 
the Fourier transform of the creation operator
requires some care.
If we define the momentum space version of
the creation operator by: 
\begin{equation}
\label{eq:spinondef}
O^{\dagger}(k_{N+1}-q_N) = \frac{1}{\sqrt{N+1}} \sum_{x = 0}^N 
e^{i(k_{N+1}-q_N)x} O^{\dagger}(x)
\end{equation}
where $k$ is an allowed momentum in the $N+1$ site
model and $q$ is an allowed momentum in the $N$ site state,
then,  the operator takes a momentum eigenstate of
the $N$ site model to an momentum eigenstate of 
the $N+1$ site model. 
(In the following we will assume periodic         %
boundary conditions, but other boundary conditions can be implemented  %
straightforwardly).                                                     %
 In particular, if $N$ is a multiple
of 4 then the groundstate has zero momentum and 
$q$ can be taken to be zero if we wish to create a single
excited spinon.
In this case,  
this construction allows us to study our approximation to the
single spinon spectral function defined by:
\begin{equation}
\label{eq:rhospinon}
A_s(k,\omega) = \sum_m |\langle m |O^{\dagger}(k_{N+1})|GS \rangle|^2
\delta(E_m - E_0-\omega)
\end{equation}
where the sum over $m$ is over a complete set of 
eigenstates of the $N+1$ site model and $|GS\rangle$ is the
groundstate of the $N$ site model.
If our conjecture were a perfect spinon creation
operator, then for the ISE model the spectral function
would be a $\delta$-function of unit weight located
on the 1 spinon dispersion relation
$\varepsilon_s(k)=\frac{1}{2}((\pi/2)^2-k^2) $. 

\begin{figure}[htb]
\psfig{file=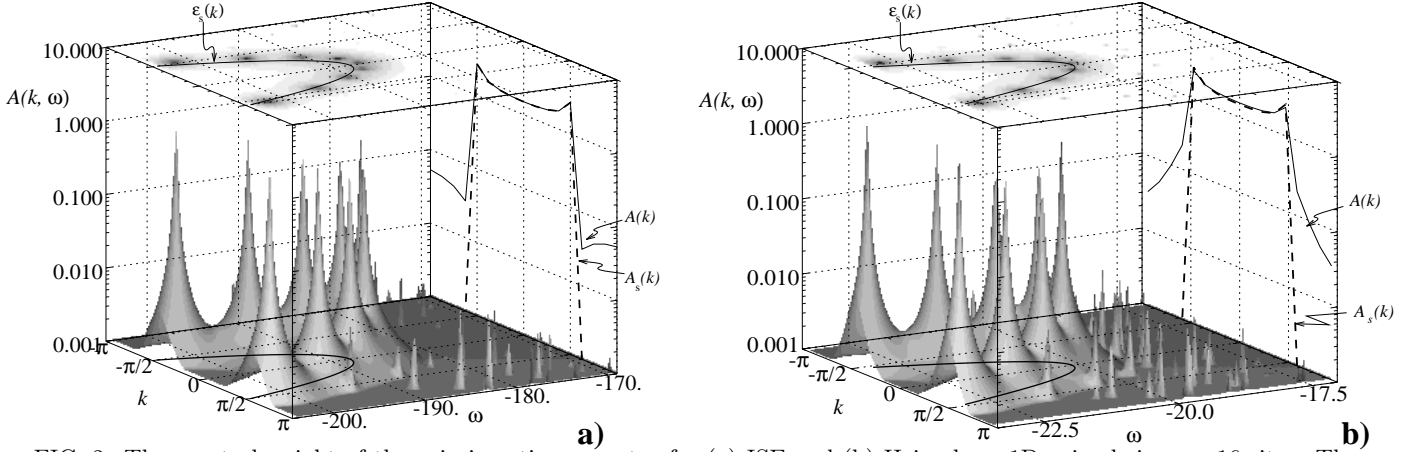,width=7.25in,angle=-90}
\widetext
\caption{The spectral weight of the spin-insertion operator for (a)
ISE and (b) Heisenberg 1D spin-chains, on 16 sites. The discrete data has been
smoothed by convolving with a Lorentzian.
Notice that the z-axis has a {\em logarithmic}-scale. 
Otherwise, on a linear scale, the
multi-spinon contributions would not be visible!
The solid line on the back wall, $A(k)$, is the $\omega$-integrated spectral
weight, or equal time spin-insertion correlator. The dashed line is
the contribution from single spinon states (the height of the large
peaks). On the top we show a
density-plot from which it is clear that virtually all the weight lies on the
1-spinon dispersion curve---$\varepsilon_s
(k)=\frac{1}{2}((\frac{\pi}{2})^2 -k^2)$ for ISE and $\pi \cos (k)$
for the Heisenberg model \protect\cite{dCP}. 
}
\narrowtext
\label{fig:rhoISE}
\end{figure}

The actual result is remarkably close to this as shown in Figure
\ref{fig:rhoISE}.  For the finite systems that we studied (up to 24 
sites), for the ISE model, at least 99.4\% of the weight of the spin
insertion seems to lie on the single spinon. However the weight in the
spectral function for $O^{\dagger}$ is not unity (as it would be for
$c^\dagger_k$ in the case of free bosons or fermions), nor is it close to
it, since we don't know {\em a priori} how to normalize our 
spinon operator
such that it creates a normalized state. However, for the ISE, using
the aforementioned almost-equivalence of spin-inserted wavefunction
and the localized spinon wavefunction, we can get an idea of what this
normalization should be like. That is, we will find the normalization
appropriate if the  
spin-insertion put {\em all} of its weight in the one-spinon state
(rather than just 99\%).
Then we need only compute the contribution to
the normalization from the  
matrix element in (\ref{eq:rhospinon}),
for the 1 spinon state (of momentum
$Q=k$).  The weight $A(k,\varepsilon_s(k))$ is given by (in an
obvious notation):
\begin{eqnarray}
\sqrt{N C^{(0)}C^{(1)}} A(k)& = &
\langle \Psi^{(1)}_{N+1}(Q= k)|\cdot\sum_{j=0}^N
|\Psi^{(1)}_{N+1}(j)\rangle e^{-ikj} \nonumber\\
&= &
\sum_{i,j=0}^N \langle \Psi_{N+1}^{(1)}(i)|\Psi_{N+1}^{(1)}(j)\rangle
e^{ik(i-j)},
\label{eq:MEFT}
\end{eqnarray}
where \mbox{$C^{(0)}=\frac{N!}{M!} N^M 2^{-M(N-1)}$} and $C^{(1)}= M!
N^M 2^{-M(N-1)}$ are the normalization of respectively states
$\Psi_{N}^{(0)}$ (eq. \ref{eq:wfgs}) and $\Psi_{N+1}^{(1)}(x)$ (eq.
\ref{onespinon}). The relevant matrix element in the RHS of this
equation, involving sums over locations of down-spins, can be
computed, after one realizes that these sums can be replaced by
integrals of the same expression, due to the polynomial nature of the
summand \cite{HS},\cite{TH94}. Methods to compute integrals of this
kind have been developed in the context of the Calogero-Sutherland
model \cite{Suth71} and random matrices, using for instance
Jack-polynomials, and matrix-model correlators\cite{Ha94,ftnote2}.  We
will follow instead Sutherland's recipe \cite{Suth71},\cite{Suth92}.
There we find, after a little bit of algebra that
($M\equiv\frac{N}{2}$):
\begin{eqnarray}
g(i-j) &\equiv &\langle \Psi_{N+1}^{(1)}(i)|\Psi_{N+1}^{(1)}(j)\rangle
=\langle\prod_{l=1}^M u(\theta_l)\rangle={\rm det}F_{p,q}\nonumber\\
 u(\theta) &=&\frac{1}{2}(\cos\left(\frac{i-j}{N}\pi\right)-\cos(\theta))
\nonumber\\ 
F_{p,q}&=&\int_{-\pi}^\pi \frac{d\theta}{2\pi p}
u(\theta)\left\{ p\cos\theta p\cos\theta q + q\sin p\theta \sin
q\theta\right\}\nonumber\\ 
&=& \frac{1}{4}\left\{ \cos\left(\pi {\scriptscriptstyle \frac{
i-j}{N}}\right)\delta_{pq}
-\frac{1}{2}\delta_{p,q+1}-\frac{1}{2}\delta_{p,q-1}\right\}
({\scriptscriptstyle 1+\frac{p}{q}})\nonumber\\
& & {\rm and\,\, } p,q=\frac{1}{2}\ldots M-\frac{1}{2}
\label{eq:detdef}
\end{eqnarray}
If we set $x=\cos\left(\frac{i-j}{N}\pi\right)$ the ensuing
(tridiagonal!) determinant observes the recursion relation:
\begin{equation}
d_{n+1}(x)=(2n+1)d_n(x)-n^2d_{n-1}(x)
\end{equation}
with initial conditions $d_0=1,\, d_1=x$. This is exactly the
definition of the Legendre Polynomials: $d_M(x)=M! P_M(x)$ (with
normalization $P_M(1)=1$). Then the Fourier transform in eq.\
(\ref{eq:MEFT}) is given by \cite{GR}:
\begin{eqnarray}
A(k) & = & 
(N+1)\left({\scriptstyle \begin{array}{c} 2r\\r\end{array}}\right)
\left({\scriptstyle\begin{array}{c} 2(M-r)\\M-r\end{array}}\right) 2^{-2M}
\nonumber\\
&\,\,\stackrel{\frac{1}{N}\ll\frac{\pi}{2}-k\ll 1}{\simeq}\,\, &
\frac{2}{\sqrt{\left(\frac{\pi}{2}\right)^2-k^2}},
\label{eq:weight}
\end{eqnarray}
and $k=\frac{\pi}{2}-\frac{2\pi}{N+1}(r+\frac{1}{4})$. In the
derivation we have chosen $M$ even to prevent cumbersome notation. See
the remark above about where the weight is located, depending on the
parity of $M$.  We notice that the weight diverges at what Haldane
calls the spinon ``pseudo-fermi surface''.  The square root
divergence is exactly that expected for spectral function of the spin
part of the electron operator obtained in Abelian bosonization; the
exponent should be 
$\frac{K_\sigma}{4}+\frac{1}{4K_\sigma}=\frac{1}{2}$, for $K_\sigma =1$,
i.e.\ isotropic spin exchange\cite{mecondmat}.

Although almost all of the effect of
the operator defined by
eq. \ref{eq:spinondef} is to create
a single spinon in an ISE models, there is some 
amplitude in the spectral function 
off the single spinon energy, so that
our operator
can not be taken to be identically the spinon
creation operator.  The connection between the
two is rather like that between the bare electron
operator and the quasiparticle operator in Fermi liquid
theory: in addition to the expected $\delta$-function there
are very small ``incoherent'' contributions coming from the fact
that our creation operator has
finite weight to create more than
one spinon.

In any case, the operator will be extremely
useful if we can demonstrate that it
has large finite overlap with the spinon creation
operator in a broader context than acting
on the ground state of the ISE model. For example,
the Heisenberg model is in the same universality
class as the ISE model and can be thought of as a model
of nearly free spinons (they have a marginally irrelevant 
interaction), and so we have displayed the spectral
function of the spinon operator acting on the
groundstate of that model together with the
ISE spectral function in Figure \ref{fig:rhoISE}.
For both models there is very little probability to excite states
with momentum, inside the ``pseudo-fermi surface'', weight is down there
by as much as 3 orders of magnitude, and,
we see that, for the Heisenberg 
model, the weight is
predominantly distributed on a 1-parameter family of states, which
carries at least 98 \% of the weight.
Notice that the single spinon feature in the spectral
function exists only over half the Brillouin
zone as expected from the Bethe Ansatz
solution of the Heisenberg model
and the solution of the ISE model\cite{HS}.
This is also consistent with the fact
that the two species of spinon
(up and down spin) should obey {\em mutual} $\frac{1}{2}$
statistics \cite{ftnote3}
so that the creation a spinon of either
type reduces the Hilbert space of the spinon
of the other type by $\frac{1}{2}$.  
This is realized here since adding one site and one
spinon to the model adds only $\frac{1}{2}$ of a state
for other spinons, because only one-half of the Brillouin zone is
accessible.  
The interchange of two spinon insertion
operators acting at different points,
conversely does not yield a phase of $\pm \pi/2$;
rather the leftmost spinon is shifted one
site. However, in the low energy limit,
the spinon lives near momentum $\pm \pi/2$
and the position shift mimics semionic
statistics.

\begin{figure}[htb]
\centerline{\psfig{file=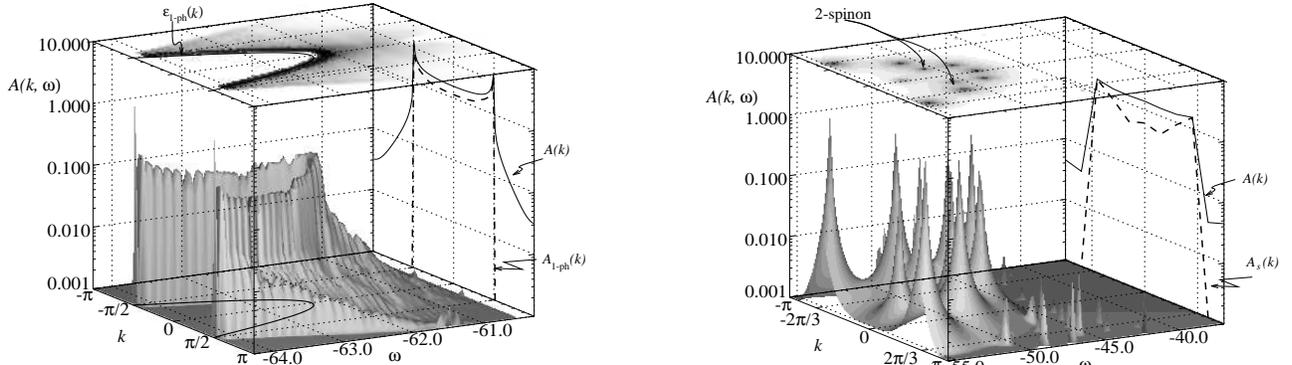,width=3in}}
\caption{The spectral weight of the spin-insertion operator for the
XY-model on 100 sites, allowing up to 3 particle hole excitations. 
The data has been smoothed by convolving with with a Lee-filter, and
plotted on a {\em log-linear} scale.
Notice that some {\em finite} weight seems to be present on the
groundstates with momenta $-\frac{\pi}{2}\leq Q\leq \frac{\pi}{2}$;
these contributions, however, vanish in the thermodynamic limit, except
for the lowest energy states close to $\pm \frac{\pi}{2}$. The latter
have measure 0, thus  restoring results from bosonization. 
The curve on top is the 1
particle-hole dispersion, other symbols are as in
fig. \protect\ref{fig:rhoISE}. 
}
\label{fig:rhoXY}
\end{figure}

Both the ISE and Heisenberg models are exactly soluble
and therefore well understood, however, our definition
can be extended to essentially any one dimensional
spin chain model, including
anisotropic spin chain model such as the
$XY$ model ($H = \sum_i S^x_i S^x_{i+1} + S^y_i S^y_{i+1}$).
This model can be mapped onto a hard-core boson
problem and then onto free fermions 
via a Jordan-Wigner transformation, so it
is readily soluble, however, the connection 
between that solution and other viewpoints
is still unclear. In particular, it should be possible
to understand  the 
$XY$ model  as model of interacting spinons.
If our spinon creation operator can really
be taken as such then we can study the underlying,
interacting 
spinons of $XY$ model and elucidate this
connection---a possibility that would not exist
were we restricted to the first quantized understanding
of the spinons available for the ISE model.
For example, we have computed the 
spinon spectral function as before for
this model with the results shown
in Figure \ref{fig:rhoXY}.

They can be compared to the
theoretical explanation based on the 
identification of the chiral parafermion,
$\exp(i \frac{1}{\sqrt{2}} \theta^{R~{\rm or}~L}_{\sigma}(x))$,
with the spinon.  In that case, the
Luttinger liquid hypothesis \cite{haldll}
for the XXZ chains predicts that the
asymptotic correlations of the spinon 
in the $XY$ model should take the form 
\begin{eqnarray}
\label{eq:xycorr}
A_{\rm spinon}(x,t) &\sim &
  \frac{e^{i \pi x /2}}{(x - vt)^{9/16}(x+vt)^\frac{1}{16}}\nonumber\\ 
&+& {\rm left~ moving~ piece} + {\rm non-universal}
\end{eqnarray}
The universal properties of the
spectral function can be computed from this
\cite{mecondmat} and it follows that:
(1) the spinon spectral function
has no $delta$-functions but rather power law singularities
at $\omega = \pm v k$, with $k$ measured from $\pm \pi/2$,
of the form: 
$|\omega - vk|^{-15/16}$ and $|\omega + vk|^{-7/16}$
and (2) the integral of the spectral function over frequency: $A(k)$
(which would simply be the $1-n_k$ for fermions or bosons)
diverges like $k^{-\frac{3}{8}}$ as $\pm \pi/2$ is approached from either
side. The ratio of the prefactors of the divergences
is
 $\frac{\sin(7\pi/16)}{\sin(\pi/16)} \sim 5.03$.
This agrees with our numerical results for this ratio
for systems of large size (about 2000 sites for $A(k)$ and 100 for
$A_{\rm spinon}(k,\omega)$).
The power law rather than $\delta$-function divergence in 
$A_{\rm spinon}$ is quite obvious when we compare
Fig. \ref{fig:rhoISE} and \ref{fig:rhoXY}: ie.\ $O^\dagger (Q)$ has
significant weight on a large number of states beyond the
lowest-energy single particle-hole state. One apparent exception to this
statement is spin-insertion with $Q\sim \pm \frac{\pi}{2}$: in that
case the larger part of the weight remains on a single particle-hole
state, even as $N\rightarrow \infty $. However, the relative weight on
this class of free Jordan-Wigner fermion states decays as
$[|\frac{\pi}{2}-Q|N ]^{-\eta} $ with $\eta\simeq 0.1$ and vanishes
(slowly) when we take $N\rightarrow \infty$ {\em before} $|Q|\rightarrow
\frac{\pi}{2} $.
Thus our spinon construction is in accord with
all of the expected properties for a spinon
creation operator in the $XY$ model.

\begin{figure}[htb]
\rule{0em}{5.7in}
\centerline{\psfig{file=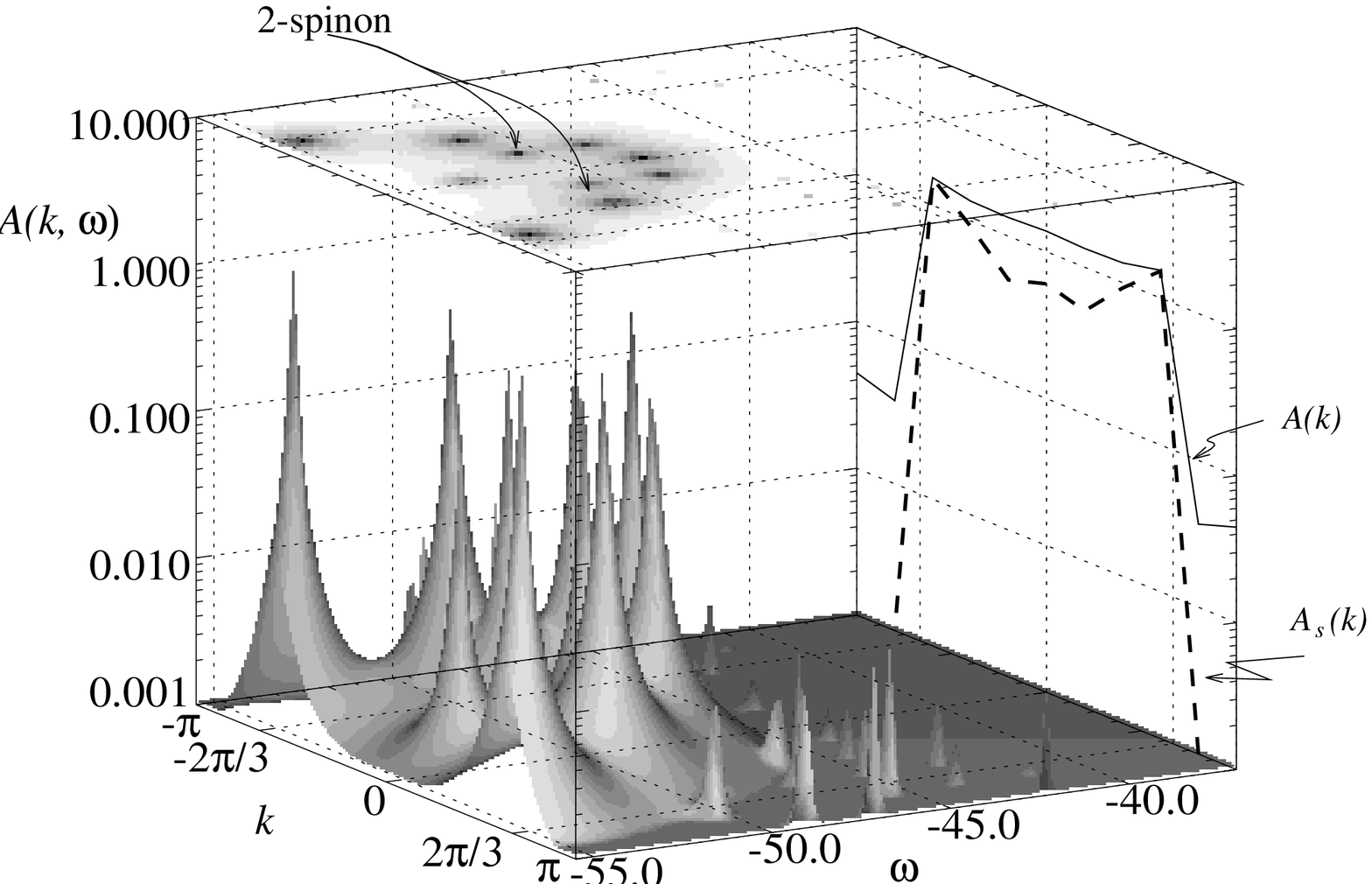,width=3in}}
\caption{The spectral weight of the spin-insertion operator for the
SU(3) Heisenberg model on 9 sites. The data has been smoothed by convolving
with a Lorentzian and plotted on a log-linear scale. The symbols are
as in fig.\ \protect\ref{fig:rhoISE}.  Notice that in addition to
the 1-spinon ('3'-representation) there is also some weight on
states with 2 anti-spinons ('$\bar{3}$'-representation).  This
feature remains presumably as $N\rightarrow\infty$ \protect\cite{motif}.}
\label{fig:rhoSU3}
\end{figure}

An analog of our spinon
operator can also be constructed for ``spin''
chains not based on $SU(2)$,
such as the $SU(3)$ generalization
of the nearest neighbor Heisenberg model in which the ``spin''
at each site is in the fundamental representation
of $SU(3)$ \cite{Suth75} (This is the model that would naturally
be obtained from an $SU(3)$ generalization of
the large $U$ Hubbard model at $\frac{1}{3}$ filling).
The results for the spectral function are shown in Fig.\ \ref{fig:rhoSU3}.

For the ISE and Heisenberg models, we knew that we were
dealing with free or weakly interacting spinons, however,
the existence of such spinons in the $SU(3)$ model,
while plausible has not been shown. The spinon  spectral function of
Fig.\ \ref{fig:rhoSU3} demonstrates that that model is
also one of nearly free spinons. Our spinon creation
operator construction is thus very useful in this context,
yielding a powerful, qualitative result quite simply.
Moreover, the spectral function we have obtained reveals
that the spinons of the $SU(3)$ model
are not semions but rather obey  mutual $\frac{1}{3}$
statistics;
that is the creation of one of any of the three 
spinon species reduces the Hilbert space 
for any of the three spinon species by $\frac{1}{3}$
of a state.
This follows from the fact that the ``spinon''
spectral function has support only over $\frac{2}{3}$
(in general $\frac{p-1}{p}$ for SU($p$) spin models)
of the Brillouin zone.
These results are in agreement with
findings for the ISE $SU(N)$ models \cite{zach} and
demonstrate the the nearest neighbor model
is almost certainly in the same universality class.

Our results for the spinon spectral function in
the Heisenberg model are directly related to the work
on the electron spectral function of the 
one dimensional Hubbard model \cite{Penc,Sorella}.
As a result of the completeness of the momentum
space spinon operators for states reached by the
insertion of a single spin, and the fact that 
our spinon creation operator creates a momentum
eigenstate, the function $C_{\sigma}(k,\omega)$
of \cite{Penc} is, in fact, identical to 
our $A_s(k,\omega)$.  Their finding that the
weight in  $C_{\sigma}(k,\omega)$ is concentrated
in half the Brillouin zone at the lowest allowed
energies is a consequence of the fact that
$C_{\sigma}(k,\omega)$
contains the single spinon spectral function plus small
corrections coming from three and higher spinon
terms.  As a result, it is dominated by the
single spinon part of the 
spectral function. This occupies only half the Brillouin
as  a consequence of the Yangian symmetry and
fractional statistics of the
spinons \cite{yangian}.  For the $XY$ model,
(also studied in \cite{Penc})
the spinons are interacting and even the single
spinon contribution to the spectral function
has weight over the entire Brillouin zone
and a power law divergence as the threshold
energy is approached from above.  This
behavior for $C_{\sigma}(k,\omega)$,
was, in fact, independently obtained in \cite{Penc}. 

In summary, we have proposed a spinon
creation operator for one dimensional
spin models and their generalizations.
We have shown that the
states created by the
proposed operator have excellent,
although not perfect, overlap with the
actual one spinon eigenstates of
the ISE model.  In addition,
for finite size Heisenberg models,
nearly all of the operator's
weight when acting on the
groundstate goes into creating an
the eigenstate with the lowest possible
energy for a given momentum---consistent
with what expects for the single
spinon creation operator given that
the Heisenberg model is a model of
spinons with marginally irrelevant
interactions.  For the $XY$ model,
the operator acting on the
groundstate creates states with a
broad distribution of energies,
but the detailed properties of the
distribution are those expected for
a spinon creation operator
(if the $XY$ model is regarded as
a Luttinger liquid model of strongly
interacting semionic spinons).
For an $SU(3)$
generalization of
the Heisenberg model, we 
find results consistent with a straightforward generalization of the
SU(2) spinon creation operator acting as a spinon creation operator
for a system of weakly interacting ``spinons''
obeying Haldane type $\frac{1}{3}$ mutual 
exclusion statistics.  This result is consistent
with expectations but, to our knowledge,
has not been demonstrated by any other means.

Together, these results indicate that
the operator proposed 
is a valid spinon creation operator,
and should be quite useful in the study
of one dimensional spin models and their
generalizations. 

One of us (S.P.S) acknowledges support from Department of
Energy grant DOE DE-FG02-90ER40542.

% now the references. delete or change fake bibitem. delete next three
%   lines and directly read in your .bbl file if you use bibtex.

\clearpage
% \begin{figure}
% \caption{}
% \label{}
% \end{figure}

\begin{references}
\bibitem{twokindsnote} There are some complications
related to that existence of two kinds of spinon as discussed
later in the text.
\bibitem{ftnote1}
The conversion from $\{ \sigma_i \}$ to $\{ n_i \}$ basis is based on the
identity $\prod_{j=1}^N (1-\frac{z}{z_j})={\rm const}$ when
$z_j=e^{\frac{2\pi i}{N} j}$ and $z$ lies on the lattice $z^N=1$.
\bibitem{ftnote2} Following the
method of \protect\cite{HZ93}, now considering correlators of
Pfaffians instead of determinants---just one spinon is created instead
of two---we obtain a result, similar to the one in that reference, but
now the final integral is not
over a manifold of $4\times 4$ matrices $M$, but $2\times 2$ matrices
instead. At the saddle point this reduces to a one parameter integration:
$\int \frac{d\lambda d\omega}{\sqrt{1-\lambda^2}}
\delta(\omega-\varepsilon_s(\lambda)) e^{i\lambda x -i\omega t} $,
cf. eq.\ (\protect\ref{eq:weight}).
\bibitem{motif} The 2 anti-spinon states with significant spectral
weight are analogous to those in the ISE-model with the simplest ``motifs''
(with the fewest '1's).
\end{references}
\end{document}